\documentclass{rsauthor}
\usepackage{graphicx}
\usepackage{natbib}
\bibpunct[, ]{(}{)}{;}{a}{}{,}
\jname{Phil. Trans. R. Soc. A}
\begin{document}

\title[SN/GRB connection]{The supernova/gamma-ray burst/jet connection}

\author{Jens Hjorth}

\address{Dark Cosmology Centre, Niels Bohr Institute, University of
Copenhagen, \\
Juliane Maries Vej 30, DK-2100 Copenhagen \O, Denmark}


\label{firstpage}

\abstract{
The observed association between supernovae and gamma-ray bursts represents a 
cornerstone in our understanding of the nature of gamma-ray bursts. The 
collapsar model \citep{1999ApJ...524..262M} provides a theoretical framework 
for this connection. A key element is the launch of a bi-polar jet (seen as 
a gamma-ray burst). The resulting hot cocoon disrupts the star while the
$^{56}$Ni produced gives rise to radioactive heating of the ejecta, seen as 
a supernova. In this discussion paper I summarise the observational status of 
the supernova/gamma-ray burst connection in the context of the `engine' 
picture of jet-driven supernovae and highlight SN 2012bz/GRB 120422A --
with its luminous supernova but intermediate high-energy luminosity -- 
as a possible transition object between low-luminosity and jet gamma-ray 
bursts. The jet channel for supernova explosions may provide new insight 
into supernova explosions in general.
}

\maketitle

\section{Introduction}

SN 1998bw \citep{1998Natur.395..670G,1998Natur.395..663K}, coincident in 
space and time with GRB 980425, remains the prototype radio-bright, 
broad-lined (BL) Type Ic supernova, against which other supernova/gamma-ray 
bursts are measured up. GRB 980425, however, was peculiar, 
with an isotropic equivalent energy release in $\gamma$-rays of only 
$E_{\rm \gamma,iso}\sim 10^{48}$ erg.  The optical lightcurve of SN 1998bw,
depicted in Fig.~1, exhibited a characteristic rise to peak of about 
$M_V=-19.2$ mag in about 16 days, similar to a Ia supernova. 
\citet{1999ApJ...524..262M} 
predicted that ``{\it all gamma-ray bursts produced by the collapsar model 
will also make supernovae like SN 1998bw}''. In this model, the progenitor
star is a Wolf--Rayet star \citep{2007ARA&A..45..177C,2012ARA&A..50..107L}, 
i.e., a massive star which has shed its envelope of hydrogen and helium, 
possibly through eruptions \citep{2006ApJ...645L..45S}.


\begin{figure}
 \includegraphics[width=1.00\columnwidth]{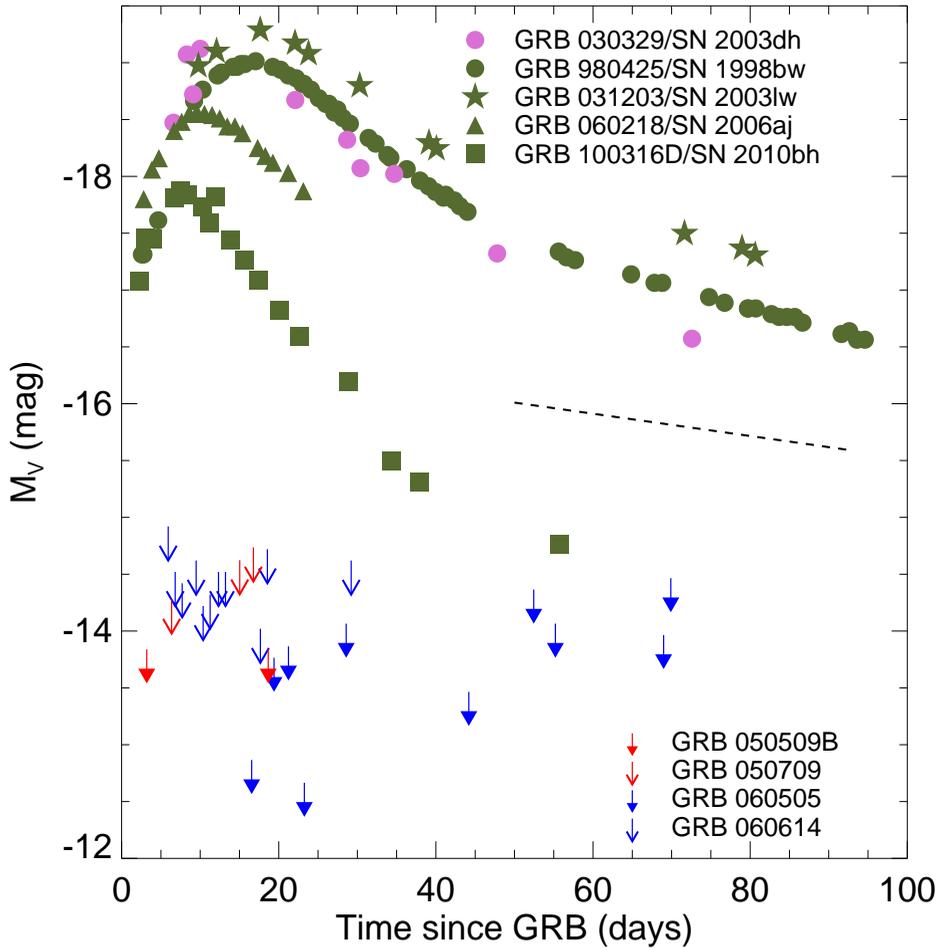}
 \caption{Optical lightcurves for the grade A \citep{2011arXiv1104.2274H}
 spectroscopic supernovae
 associated with gamma-ray bursts (excluding SN 2012bz). 
 The olive points are supernovae from
 low-luminosity gamma-ray bursts while the orchid data
 points are for SN 2003dh, associated with the jet 
 gamma-ray burst GRB 030329. There is considerable diversity in the light
 curves, regarding time to peak and peak magnitude. The $^{56}$Co decay
 slope is shown for reference (dashed line). Also shown are upper
 limits on supernova emission from long gamma-ray bursts (blue) and short
 gamma-ray bursts (red)
 \citep[adapted from][]{2011arXiv1104.2274H}.
A recent compilation of lightcurves of other supernovae associated with 
gamma-ray bursts is available in \citet{2011ApJ...740...41C}. 
          }
 \label{fig:lightcurves}
\end{figure}

The ultimate proof of a SN 1998bw-like supernova associated with a `normal'
cosmological gamma-ray burst with $E_{\rm \gamma,iso}\sim 10^{52}$ erg
came with the spectroscopic 
identification of SN 2003dh associated with GRB 030329, as a supernova 
spatially and temporally coincident with the gamma-ray burst, and with 
lightcurve properties and spectroscopic broad-line evolution very similar to
that of SN 1998bw \citep{2003ApJ...591L..17S,2003Natur.423..847H}.

The night before the Royal Society Discussion Meeting on ``New
windows on transients across the universe'', GRB 120422A was observed by
{\it Swift} and subsequently by ground-based telescopes at a redshift of 0.28.
An accompanying supernova was predicted at the meeting and indeed 
reported soon after as SN 2012bz 
\citep{2012GCN..13276...1W,2012GCN..13277...1M}.

In this paper I focus on `jet-driven' supernovae and their relation 
(or lack thereof) to the different classes of gamma-ray bursts and
highlight the importance of SN 2012bz/GRB 120422A. More comprehensive 
(and less speculative) reviews of the supernova/gamma-ray burst connection 
can be found in \citet{2006ARA&A..44..507W} and \citet{2011arXiv1104.2274H}.

\section{Supernovae associated with long-duration gamma-ray bursts}

There seem to be two types of long-duration gamma-ray bursts. The
exact division is unclear but we will discuss them in turn below.

\subsection{Low-luminosity gamma-ray bursts}

Low-luminosity gamma-ray bursts (also termed `sub-energetic' or
`nearby' bursts) seem
to be about 100 times as common as the other class discussed
below \citep{2006Natur.442.1011P}, but because of their low 
luminosities they are primarily found at low redshifts as
rare events (one every $\sim 3$ years). They typically 
have single-peak high-energy prompt lightcurves, soft high-energy 
spectra, and are often found to be X-ray flashes, i.e.,
gamma-ray bursts with peak energies below $\sim 50$ keV.
Observational evidence suggests that the radio and high-energy
emission is due to the breakout of a relativistic shock
from the surrounding massive wind of the progenitor star
\citep{1968CaJPh..46..476C,1998Natur.395..663K,2006Natur.442.1008C,
2006Natur.442.1014S,2012ApJ...747...88N}. Apart from SN 2003dh (and 
possibly SN 2012bz), the best studied supernovae related to gamma-ray bursts 
are all members of this class. Their lightcurves are shown in Fig.~1.

\subsection{Jet gamma-ray bursts}

\begin{figure}
 \includegraphics[width=1.00\columnwidth,clip=]{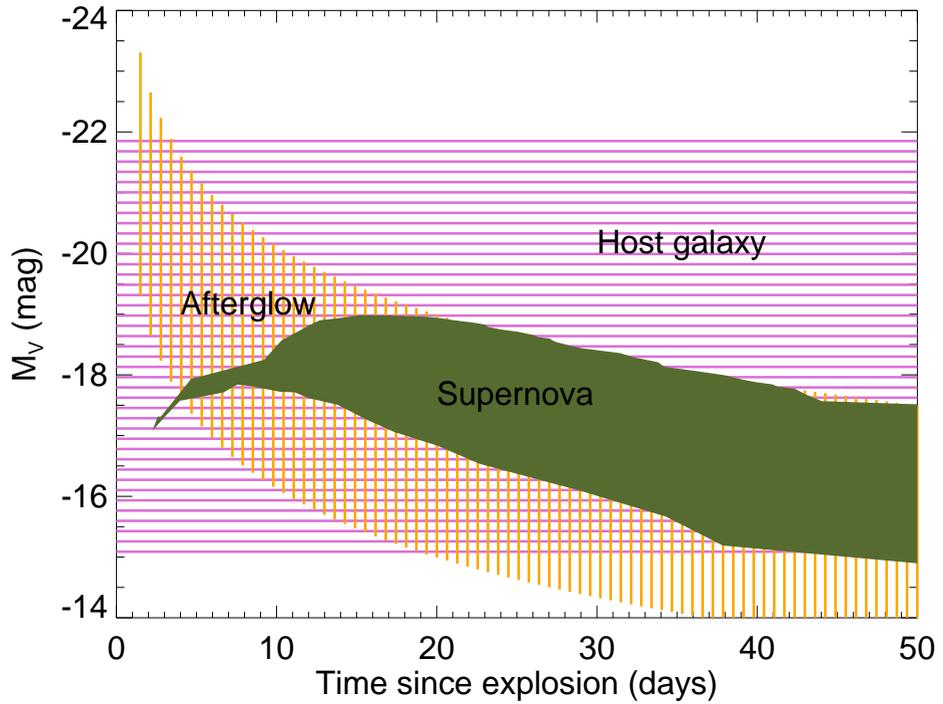}
 \caption{Schematic diagram illustrating the challenges in detecting supernova 
 light on the background of the gamma-ray burst afterglow and the host galaxy. 
 The supernova region
 (olive) reflects the range in lightcurves shown in Fig.~1. The 
 afterglow is assumed to decay as $t^{-1.5}$; the afterglow region 
 reflects the range in 
 afterglow brightness reported by \citet{2010ApJ...720.1513K}. The range in 
 host galaxy magnitudes reflect those
 detected in the TOUGH survey \citep{2012ApJ...756..187H}. 
 The diagram shows that observed lightcurves 
 may either be supernova, afterglow or host galaxy dominated. All situations
 are encountered in nature.
 }
 \label{fig:compare}
\end{figure}

These are also known as `normal' or `cosmological' gamma-ray bursts 
(or `collapsar' bursts, although this is a somewhat theory-laden term) 
and are characterised by more complex prompt emission lightcurves and
higher energies, luminosities and peak energies. They are believed to arise 
from emission from a relativistic jet at large distances from the progenitor 
star.

Observing a supernova related to a gamma-ray burst at higher redshift
is challenging because of possible contamination by the host galaxy
(which often appears unresolved in ground-based observations) and 
the afterglow. This is illustrated
in Fig.~2. Indeed, as shown by \citet{2004ApJ...606..381L}, the
lightcurve of GRB 0303029 did not exhibit a conspicuous lightcurve
bump from SN 2003dh because it was afterglow dominated. 
Besides 2003dh
(shown in Fig.~1), the best example of a supernova related to a gamma-ray 
burst in this class is SN 2010ma \citep{2011ApJ...735L..24S}
(and possibly SN 2012bz).
\subsection{Statistical properties of supernovae associated with 
gamma-ray bursts}

Inferring statistical properties of supernovae associated with
gamma-ray bursts requires a well-defined sample.
For this purpose \citet{2011arXiv1104.2274H} devised a grading
scheme for each supernova claimed in the literature to be related 
to a gamma-ray burst. The evidence for a supernova was graded A--E.%
\footnote{We have created a website (http://www.dark-cosmology.dk/GRBSN) 
dedicated to providing updates to the list of supernovae related to gamma-ray 
bursts, the grading of the observational evidence for a supernova, and 
supplementary 
information on the supernovae and the associated gamma-ray bursts.}
Based on supernovae with grades A,B,C we plot in Fig.~3 the
distribution of peak supernova magnitudes as a function of
isotropic equivalent luminosity in $\gamma$-rays. Defining
$L_{\rm \gamma,iso} = E_{\rm \gamma,iso} T_{90} ^{-1} (1+z)$
we have tentatively identified low-luminosity gamma-ray bursts as having
$L_{\rm \gamma,iso} < 10^{48.5}$ erg s$^{-1}$ and jet gamma-ray bursts as
having $L_{\rm \gamma,iso} > 10^{49.5}$ erg s$^{-1}$. 
There is a real dispersion in the peak magnitudes; supernovae related
to gamma-ray bursts are evidently not standard candles. It remains an 
open question whether they are standardizable similar to Type Ia supernovae
\citep{2005ApJ...626L...5S,2011ApJ...740...41C}. It is evident that the
lightcurves of the subsample of supernovae shown in Fig.~1 exhibit a clear 
correlation between the peak magnitude and the width of the peak.

We note that beaming and viewing angle can significantly affect the 
inferred high-energy luminosity \citep{2010arXiv1012.5101G}. For example,
GRB 091127, which appears in the high-luminosity (jet) part of 
Fig.~3, has been suggested to be a sub-energetic burst 
\citep{2012ApJ...761...50T} due to a beaming correction. 
Nevertheless, it is evident that there appears to be a
parabola-shaped upper envelope to the brightness of supernovae
as a function of high-energy luminosity. By the time of 
the meeting there were no gamma-ray bursts with convincing supernovae in 
the range $10^{48.5}$ erg $<E_{\rm \gamma,iso} < 10^{49.5}$ erg.
This changed with SN 2012bz/GRB 120422A which fills this gap in
high-energy luminosity as one of the brightest supernovae associated
with a gamma-ray burst ever detected \citep{2012A&A...547A..82M}.

\begin{figure}
 \includegraphics[width=1.00\columnwidth,clip=]{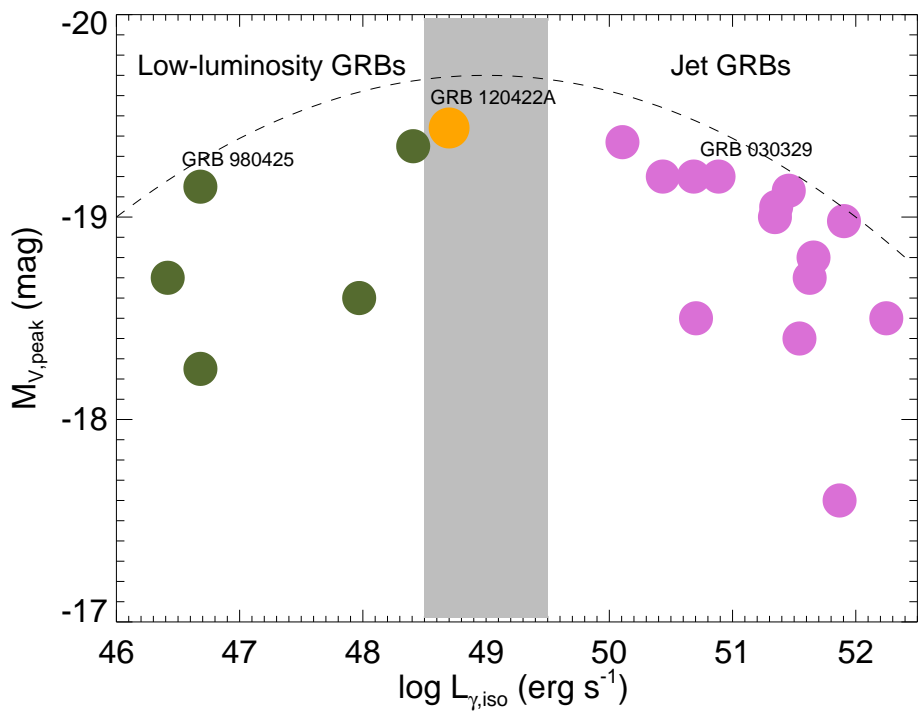}
 \caption{Supernova optical peak brightness versus gamma-ray burst isotropic
 luminosity \citep{2008MNRAS.391..577A,2009A&A...508..173A} for grade A,B,C 
 systems \citep{2011arXiv1104.2274H}. Low-luminosity gamma-ray burst supernovae 
 ($E_{\rm \gamma,iso} < 10^{48.5}$ erg, olive) and supernovae from 
 jet gamma-ray bursts ($E_{\rm \gamma,iso} > 10^{49.5}$ erg, orchid) have 
 similar distributions of peak brightness. SN 2012bz/GRB 120422A is 
 highlighted (orange) as a possible transition object in the grey area
 $10^{48.5}$ erg $<E_{\rm \gamma,iso} < 10^{49.5}$ erg
 \citep{2012ApJ...756..190Z}.
 }
 \label{fig:emv}
\end{figure}

How do these peak magnitudes compare to other similar supernovae,
i.e., Type Ic supernovae, with no hydrogen or helium in their
spectra? Using the well-defined sample of normal Ic supernovae
from \citet{2011ApJ...741...97D} and a more heterogeneous sample of 
broad-lined Ic supernovae from a variety of sources, we plot in
Fig.~4 cumulative histograms of their peak magnitudes. Type Ic supernovae
seem to be fainter than supernovae related to gamma-ray bursts
while the situation is less clear-cut for Ic-BL supernovae with
no GRBs. We note
that strong observational evidence (grade A--C) quite naturally
will bias our sample against fainter supernovae.

\begin{figure}
 \includegraphics[width=1.00\columnwidth,clip=]{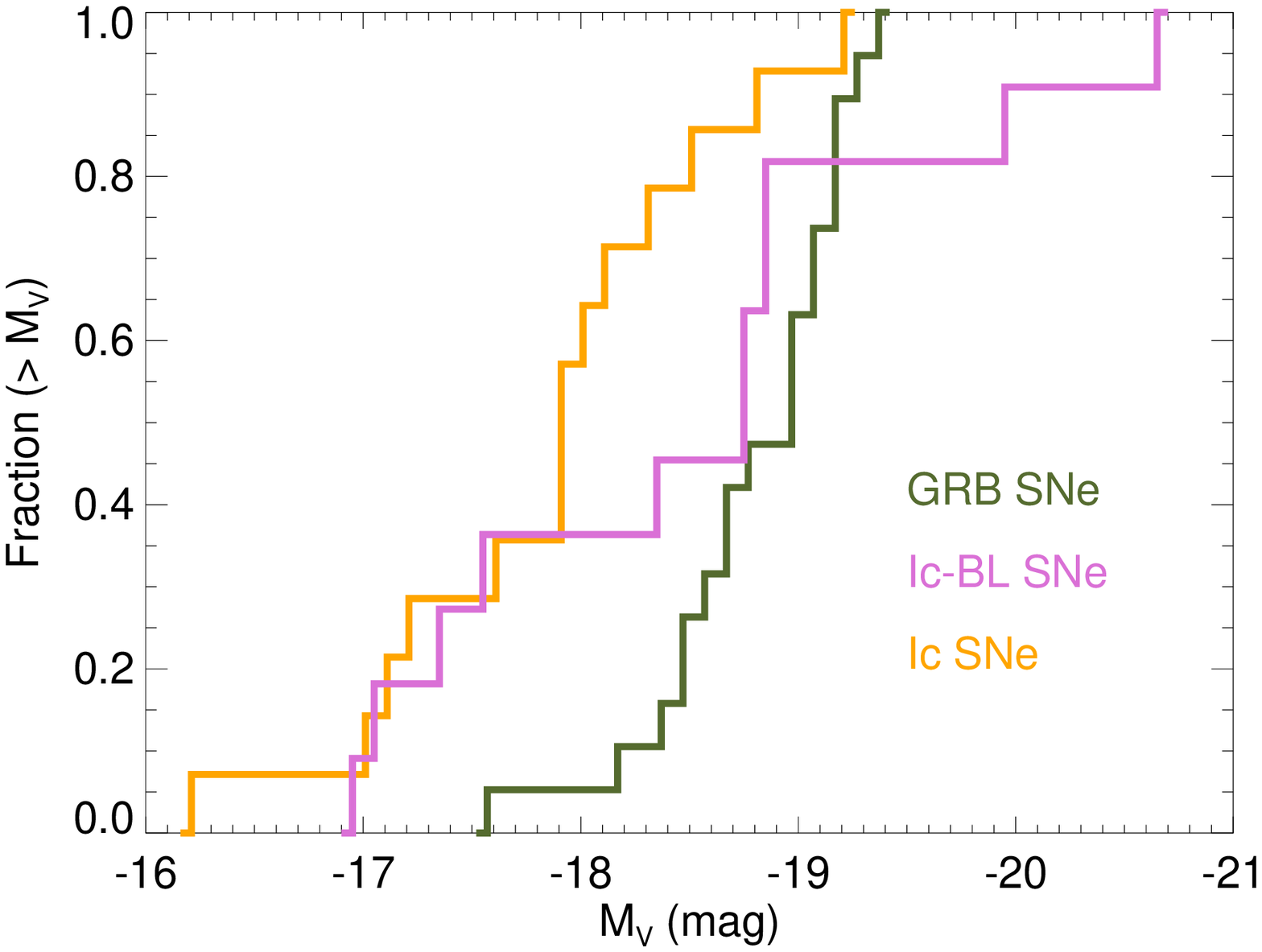}
 \caption{Cumulative distributions of the brightness of different
 kinds of Ic supernovae. The supernovae associated with gamma-ray
 bursts graded A, B, or C are from \citet{2011arXiv1104.2274H}. 
 The normal Ic supernovae are from \citet{2011ApJ...741...97D}.
 The Ic-BL distribution comes from a variety of sources and as such
 represents a more ill-defined sample. The gamma-ray burst supernovae
 generally appear brighter than normal Ic supernovae, although it 
 should be noted that they are likely biased against faint systems.
 The brightness distribution of Ic-BL is probably consistent with that 
 of gamma-ray burst supernovae although there may be a lack of very 
 bright gamma-ray burst supernovae. 
 }
 \label{fig:histogram}
\end{figure}

The comparison to Ic-BL is 
interesting because the rates of low-luminosity gamma-ray bursts and 
Ic-BL are comparable, suggesting perhaps a common origin and indicates
that low-luminosity gamma-ray bursts, as expected, may not be strongly 
beamed \citep{2004ApJ...607L..17P}. 

\section{Supernova-less gamma-ray bursts}

Two classes of gamma-ray bursts are not accompanied by bright supernovae.

\subsection{Short gamma-ray bursts}

Short-duration, hard-spectrum gamma-ray bursts \citep{1993ApJ...413L.101K},
with durations $T_{90} < 2$~s, are known not to lead to supernovae. In Fig.~1 
we have plotted the upper limits on the existence of supernovae 
accompanying GRBs 050509B \citep{2005ApJ...630L.117H} and 050709 
\citep{2005Natur.437..859H}. These constrain any supernova to be about 
100 times fainter than SN 1998bw at peak. This is consistent with short 
gamma-ray bursts being the results of compact object mergers.

The data also rule out the existence of an early rebrightning in GRB 050509B 
\citep{2005ApJ...630L.117H} at 1.5 days in the restframe. Bright transient 
emission, dubbed a `mini SN' \citep{1998ApJ...507L..59L,2002MNRAS.336L...7R},
`kilonova' \citep{2010MNRAS.406.2650M} or `macronova' 
\citep{2005astro.ph.10256K}, is expected to peak around the optical-UV range 
within a day or so with a semi-thermal spectrum \citep{1998ApJ...507L..59L}. 
GRB 050509B sets very strong contraints on such emission
\citep{2005ApJ...630L.117H,2010MNRAS.404..963K,2011ApJ...736L..21R}.

\subsection{Long supernova-less gamma-ray bursts}

Perhaps surprisingly, some long-duration gamma-ray bursts are not
accompanied by bright supernovae. As shown in Fig.~1, the constraints
on GRBs 060505 and 060614 are about as constraining as the those 
related to the short gamma-ray bursts discussed above. These puzzling 
systems may be related to non-$^{56}$Ni producing supernovae or they may be
merger gamma-ray bursts with longer durations than usually found
\citep{2006Natur.444.1044G, 2006Natur.444.1047F, 2006Natur.444.1050D,
2006Natur.444.1053G, 2007ApJ...662.1129O}.

\subsection{Mind the gap}

It is quite remarkable that current observations reveal a clear gap 
between the brightnesses of gamma-ray burst supernovae, at around absolute 
magnitude $-17$ to $-19$, and the upper limits on long supernova-less
gamma-ray bursts at around $-12$ to $-14$ mag. Finding faint supernovae is of
course difficult and fainter supernovae will likely be detected but the 
current factor of 100 may indicate that there is not a simple continuum
of events.

\section{Engine-driven supernovae}

The collapsar model \citep{1999ApJ...524..262M} 
operates with two time scales, the duration of the
active `engine' (jet), $t_E$, and the time for shock breakout, $t_S$. 
A successful gamma-ray burst requires the engine to be active for longer 
than the shock-breakout time. 
\citet{2011ApJ...739L..55B} and \citet{2012ApJ...750...68L} have used
this picture to explore the consequences of the relative durations for
the resulting supernovae and gamma-ray bursts
\citep[an alternative jet scenario is presented by][]{2011MNRAS.416.1697P}:

\begin{itemize}
\item
$t_E > t_S$: a normal jet gamma-ray burst accompanied by a Ic-BL is produced
\item
$t_E \approx t_S$: a low-luminosity gamma-ray burst accompanied by a Ic-BL 
or a relativistic Ic-BL with no gamma-ray burst is produced
\item
$t_E < t_S$: a non-relativistic supernova but no gamma-ray burst is produced
\end{itemize}
In this picture, relativistic supernovae, like SN 2009bb
\citep{2010Natur.463..513S,2011ApJ...728...14P} are jet-driven supernovae,
similar to low-luminosity gamma-ray bursts. It is worth noting
that a low luminosity is not necessarily synomynous with a short
engine duration, i.e., it may be possible to have low-luminosity
jet gamma-ray bursts, such as possibly GRB 120422A.

In the collapsar model, one could also imagine that the engine does not 
occur in a stripped-envelope core-supernova but in a Type II supernova with a 
hydrogen and/or helium layer which would prevent the escape of the jet 
\citep[see also][]{2003ApJ...591..288H}. Such massive stars 
may have a dense circumstellar medium which would make them appear as 
Type IIn supernovae, as suggested by e.g., \citet{2003astro.ph..8136N} and
\citet{2012ApJ...752L...2C}. Recently a possible jet-powered IIn (SN 2010jp)
was reported \citep{2012MNRAS.420.1135S}, albeit not a relativistic one.

The picture regarding jet-driven supernovae and gamma-ray bursts that emerges 
from the discussion in this paper is summarised in Table~1. 
SN 2012bz/GRB 120422A, which may be a transition object between the 
low-luminosity and jet gamma-ray bursts, reminds us that this fairly simple
picture could easily be more complex.

 \begin{table}[h]
  \caption{The supernova/gamma-ray burst/jet connection}
    \begin{tabular}{lll}
     \hline \hline
Core-collapse supernovae & Supernova/gamma-ray bursts    & Gamma-ray bursts \\
     \hline
Relativistic Ic-BL  & Low-luminosity GRBs     & Fall-back supernovae? \\
(SN 2009bb)         & (SN 1998bw/GRB 980425)  & (GRB 060505)          \\
                    &                         &                       \\
Type IIn?           & Jet GRBs                & Mergers               \\
(SN 2010jp)         & (SN 2003dh/GRB 030329)  & (GRB 050509B)         \\
     \hline \hline
    \end{tabular}
  \label{t:table}
\end{table}

\ack{
I acknowledge discussions with Andrew MacFadyen, Paul Crowther, 
Davide Lazzati, Enrico Ramirez-Ruiz, Darach Watson and Tsvi Piran.
I thank the anonymous referees for helpful comments.
The Dark Cosmology Centre is funded by the Danish National Research
Foundation. 
}

\bibliography{hjorth}

\begin{thebibliography}{54}
\expandafter\ifx\csname natexlab\endcsname\relax\def\natexlab#1{#1}\fi

\bibitem[{{Amati} {et~al.}(2009){Amati}, {Frontera}, \&
  {Guidorzi}}]{2009A&A...508..173A}
{Amati}, L., {Frontera}, F., \& {Guidorzi}, C. 2009, \aap, 508, 173

\bibitem[{{Amati} {et~al.}(2008){Amati}, {Guidorzi}, {Frontera}, {Della Valle},
  {Finelli}, {Landi}, \& {Montanari}}]{2008MNRAS.391..577A}
{Amati}, L., {Guidorzi}, C., {Frontera}, F., {Della Valle}, M., {Finelli}, F.,
  {Landi}, R., \& {Montanari}, E. 2008, \mnras, 391, 577

\bibitem[{{Bromberg} {et~al.}(2011){Bromberg}, {Nakar}, \&
  {Piran}}]{2011ApJ...739L..55B}
{Bromberg}, O., {Nakar}, E., \& {Piran}, T. 2011, \apjl, 739, L55

\bibitem[{{Campana} {et~al.}(2006){Campana}, {Mangano}, {Blustin}, {Brown},
  {Burrows}, {Chincarini}, {Cummings}, {Cusumano}, {Della Valle}, {Malesani},
  {M{\'e}sz{\'a}ros}, {Nousek}, {Page}, {Sakamoto}, {Waxman}, {Zhang}, {Dai},
  {Gehrels}, {Immler}, {Marshall}, {Mason}, {Moretti}, {O'Brien}, {Osborne},
  {Page}, {Romano}, {Roming}, {Tagliaferri}, {Cominsky}, {Giommi}, {Godet},
  {Kennea}, {Krimm}, {Angelini}, {Barthelmy}, {Boyd}, {Palmer}, {Wells}, \&
  {White}}]{2006Natur.442.1008C}
{Campana}, S., {et~al.} 2006, \nat, 442, 1008

\bibitem[{{Cano} {et~al.}(2011){Cano}, {Bersier}, {Guidorzi}, {Kobayashi},
  {Levan}, {Tanvir}, {Wiersema}, {D'Avanzo}, {Fruchter}, {Garnavich}, {Gomboc},
  {Gorosabel}, {Kasen}, {Kopa{\v c}}, {Margutti}, {Mazzali}, {Melandri},
  {Mundell}, {Nugent}, {Pian}, {Smith}, {Steele}, {Wijers}, \&
  {Woosley}}]{2011ApJ...740...41C}
{Cano}, Z., {et~al.} 2011, \apj, 740, 41

\bibitem[{{Chevalier}(2012)}]{2012ApJ...752L...2C}
{Chevalier}, R.~A. 2012, \apjl, 752, L2

\bibitem[{{Colgate}(1968)}]{1968CaJPh..46..476C}
{Colgate}, S.~A. 1968, Canadian Journal of Physics, 46, 476

\bibitem[{{Crowther}(2007)}]{2007ARA&A..45..177C}
{Crowther}, P.~A. 2007, \araa, 45, 177

\bibitem[{{Della Valle} {et~al.}(2006){Della Valle}, {Chincarini}, {Panagia},
  {Tagliaferri}, {Malesani}, {Testa}, {Fugazza}, {Campana}, {Covino},
  {Mangano}, {Antonelli}, {D'Avanzo}, {Hurley}, {Mirabel}, {Pellizza},
  {Piranomonte}, \& {Stella}}]{2006Natur.444.1050D}
{Della Valle}, M., {et~al.} 2006, \nat, 444, 1050

\bibitem[{{Drout} {et~al.}(2011){Drout}, {Soderberg}, {Gal-Yam}, {Cenko},
  {Fox}, {Leonard}, {Sand}, {Moon}, {Arcavi}, \& {Green}}]{2011ApJ...741...97D}
{Drout}, M.~R., {et~al.} 2011, \apj, 741, 97

\bibitem[{{Fynbo} {et~al.}(2006){Fynbo}, {Watson}, {Th{\"o}ne}, {Sollerman},
  {Bloom}, {Davis}, {Hjorth}, {Jakobsson}, {J{\o}rgensen}, {Graham},
  {Fruchter}, {Bersier}, {Kewley}, {Cassan}, {Castro Cer{\'o}n}, {Foley},
  {Gorosabel}, {Hinse}, {Horne}, {Jensen}, {Klose}, {Kocevski}, {Marquette},
  {Perley}, {Ramirez-Ruiz}, {Stritzinger}, {Vreeswijk}, {Wijers}, {Woller},
  {Xu}, \& {Zub}}]{2006Natur.444.1047F}
{Fynbo}, J.~P.~U., {et~al.} 2006, \nat, 444, 1047

\bibitem[{{Gal-Yam} {et~al.}(2006){Gal-Yam}, {Fox}, {Price}, {Ofek}, {Davis},
  {Leonard}, {Soderberg}, {Schmidt}, {Lewis}, {Peterson}, {Kulkarni}, {Berger},
  {Cenko}, {Sari}, {Sharon}, {Frail}, {Moon}, {Brown}, {Cucchiara}, {Harrison},
  {Piran}, {Persson}, {McCarthy}, {Penprase}, {Chevalier}, \&
  {MacFadyen}}]{2006Natur.444.1053G}
{Gal-Yam}, A., {et~al.} 2006, \nat, 444, 1053

\bibitem[{{Galama} {et~al.}(1998){Galama}, {Vreeswijk}, {van Paradijs},
  {Kouveliotou}, {Augusteijn}, {B{\"o}hnhardt}, {Brewer}, {Doublier},
  {Gonzalez}, {Leibundgut}, {Lidman}, {Hainaut}, {Patat}, {Heise}, {in't Zand},
  {Hurley}, {Groot}, {Strom}, {Mazzali}, {Iwamoto}, {Nomoto}, {Umeda},
  {Nakamura}, {Young}, {Suzuki}, {Shigeyama}, {Koshut}, {Kippen}, {Robinson},
  {de Wildt}, {Wijers}, {Tanvir}, {Greiner}, {Pian}, {Palazzi}, {Frontera},
  {Masetti}, {Nicastro}, {Feroci}, {Costa}, {Piro}, {Peterson}, {Tinney},
  {Boyle}, {Cannon}, {Stathakis}, {Sadler}, {Begam}, \&
  {Ianna}}]{1998Natur.395..670G}
{Galama}, T.~J., {et~al.} 1998, \nat, 395, 670

\bibitem[{{Gehrels} {et~al.}(2006){Gehrels}, {Norris}, {Barthelmy}, {Granot},
  {Kaneko}, {Kouveliotou}, {Markwardt}, {M{\'e}sz{\'a}ros}, {Nakar}, {Nousek},
  {O'Brien}, {Page}, {Palmer}, {Parsons}, {Roming}, {Sakamoto}, {Sarazin},
  {Schady}, {Stamatikos}, \& {Woosley}}]{2006Natur.444.1044G}
{Gehrels}, N., {et~al.} 2006, \nat, 444, 1044

\bibitem[{{Granot} \& {Ramirez-Ruiz}(2010)}]{2010arXiv1012.5101G}
{Granot}, J., \& {Ramirez-Ruiz}, E. 2010, arXiv:1012.5101

\bibitem[{{Heger} {et~al.}(2003){Heger}, {Fryer}, {Woosley}, {Langer}, \&
  {Hartmann}}]{2003ApJ...591..288H}
{Heger}, A., {Fryer}, C.~L., {Woosley}, S.~E., {Langer}, N., \& {Hartmann},
  D.~H. 2003, \apj, 591, 288

\bibitem[{{Hjorth} \& {Bloom}(2011)}]{2011arXiv1104.2274H}
{Hjorth}, J., \& {Bloom}, J.~S. 2011, arXiv:1104.2274

\bibitem[{{Hjorth} {et~al.}(2003){Hjorth}, {Sollerman}, {M{\o}ller}, {Fynbo},
  {Woosley}, {Kouveliotou}, {Tanvir}, {Greiner}, {Andersen}, {Castro-Tirado},
  {Castro Cer{\'o}n}, {Fruchter}, {Gorosabel}, {Jakobsson}, {Kaper}, {Klose},
  {Masetti}, {Pedersen}, {Pedersen}, {Pian}, {Palazzi}, {Rhoads}, {Rol}, {van
  den Heuvel}, {Vreeswijk}, {Watson}, \& {Wijers}}]{2003Natur.423..847H}
{Hjorth}, J., {et~al.} 2003, \nat, 423, 847

\bibitem[{{Hjorth} {et~al.}(2005{\natexlab{a}}){Hjorth}, {Sollerman},
  {Gorosabel}, {Granot}, {Klose}, {Kouveliotou}, {Melinder}, {Ramirez-Ruiz},
  {Starling}, {Thomsen}, {Andersen}, {Fynbo}, {Jensen}, {Vreeswijk}, {Castro
  Cer{\'o}n}, {Jakobsson}, {Levan}, {Pedersen}, {Rhoads}, {Tanvir}, {Watson},
  \& {Wijers}}]{2005ApJ...630L.117H}
---. 2005{\natexlab{a}}, \apjl, 630, L117

\bibitem[{{Hjorth} {et~al.}(2005{\natexlab{b}}){Hjorth}, {Watson}, {Fynbo},
  {Price}, {Jensen}, {J{\o}rgensen}, {Kubas}, {Gorosabel}, {Jakobsson},
  {Sollerman}, {Pedersen}, \& {Kouveliotou}}]{2005Natur.437..859H}
---. 2005{\natexlab{b}}, \nat, 437, 859

\bibitem[{{Hjorth} {et~al.}(2012){Hjorth}, {Malesani}, {Jakobsson}, {Jaunsen},
  {Fynbo}, {Gorosabel}, {Kr{\"u}hler}, {Levan}, {Micha{\l}owski},
  {Milvang-Jensen}, {M{\o}ller}, {Schulze}, {Tanvir}, \&
  {Watson}}]{2012ApJ...756..187H}
---. 2012, \apj, 756, 187

\bibitem[{{Kann} {et~al.}(2010){Kann}, {Klose}, {Zhang}, {Malesani}, {Nakar},
  {Pozanenko}, {Wilson}, {Butler}, {Jakobsson}, {Schulze}, {Andreev},
  {Antonelli}, {Bikmaev}, {Biryukov}, {B{\"o}ttcher}, {Burenin}, {Castro
  Cer{\'o}n}, {Castro-Tirado}, {Chincarini}, {Cobb}, {Covino}, {D'Avanzo},
  {D'Elia}, {Della Valle}, {de Ugarte Postigo}, {Efimov}, {Ferrero}, {Fugazza},
  {Fynbo}, {G{\aa}lfalk}, {Grundahl}, {Gorosabel}, {Gupta}, {Guziy}, {Hafizov},
  {Hjorth}, {Holhjem}, {Ibrahimov}, {Im}, {Israel}, {Je{\'l}inek}, {Jensen},
  {Karimov}, {Khamitov}, {Kizilo{\v g}lu}, {Klunko}, {Kub{\'a}nek}, {Kutyrev},
  {Laursen}, {Levan}, {Mannucci}, {Martin}, {Mescheryakov}, {Mirabal},
  {Norris}, {Ovaldsen}, {Paraficz}, {Pavlenko}, {Piranomonte}, {Rossi},
  {Rumyantsev}, {Salinas}, {Sergeev}, {Sharapov}, {Sollerman}, {Stecklum},
  {Stella}, {Tagliaferri}, {Tanvir}, {Telting}, {Testa}, {Updike}, {Volnova},
  {Watson}, {Wiersema}, \& {Xu}}]{2010ApJ...720.1513K}
{Kann}, D.~A., {et~al.} 2010, \apj, 720, 1513

\bibitem[{{Kocevski} {et~al.}(2010){Kocevski}, {Th{\"o}ne}, {Ramirez-Ruiz},
  {Bloom}, {Granot}, {Butler}, {Perley}, {Modjaz}, {Lee}, {Cobb}, {Levan},
  {Tanvir}, \& {Covino}}]{2010MNRAS.404..963K}
{Kocevski}, D., {et~al.} 2010, \mnras, 404, 963

\bibitem[{{Kouveliotou} {et~al.}(1993){Kouveliotou}, {Meegan}, {Fishman},
  {Bhat}, {Briggs}, {Koshut}, {Paciesas}, \& {Pendleton}}]{1993ApJ...413L.101K}
{Kouveliotou}, C., {Meegan}, C.~A., {Fishman}, G.~J., {Bhat}, N.~P., {Briggs},
  M.~S., {Koshut}, T.~M., {Paciesas}, W.~S., \& {Pendleton}, G.~N. 1993, \apjl,
  413, L101

\bibitem[{{Kulkarni}(2005)}]{2005astro.ph.10256K}
{Kulkarni}, S.~R. 2005, arXiv:astro-ph/0510256

\bibitem[{{Kulkarni} {et~al.}(1998){Kulkarni}, {Frail}, {Wieringa}, {Ekers},
  {Sadler}, {Wark}, {Higdon}, {Phinney}, \& {Bloom}}]{1998Natur.395..663K}
{Kulkarni}, S.~R., {et~al.} 1998, \nat, 395, 663

\bibitem[{{Langer}(2012)}]{2012ARA&A..50..107L}
{Langer}, N. 2012, \araa, 50, 107

\bibitem[{{Lazzati} {et~al.}(2012){Lazzati}, {Morsony}, {Blackwell}, \&
  {Begelman}}]{2012ApJ...750...68L}
{Lazzati}, D., {Morsony}, B.~J., {Blackwell}, C.~H., \& {Begelman}, M.~C. 2012,
  \apj, 750, 68

\bibitem[{{Li} \& {Paczy{\'n}ski}(1998)}]{1998ApJ...507L..59L}
{Li}, L.-X., \& {Paczy{\'n}ski}, B. 1998, \apjl, 507, L59

\bibitem[{{Lipkin} {et~al.}(2004){Lipkin}, {Ofek}, {Gal-Yam}, {Leibowitz},
  {Poznanski}, {Kaspi}, {Polishook}, {Kulkarni}, {Fox}, {Berger}, {Mirabal},
  {Halpern}, {Bureau}, {Fathi}, {Price}, {Peterson}, {Frebel}, {Schmidt},
  {Orosz}, {Fitzgerald}, {Bloom}, {van Dokkum}, {Bailyn}, {Buxton}, \&
  {Barsony}}]{2004ApJ...606..381L}
{Lipkin}, Y.~M., {et~al.} 2004, \apj, 606, 381

\bibitem[{{MacFadyen} \& {Woosley}(1999)}]{1999ApJ...524..262M}
{MacFadyen}, A.~I., \& {Woosley}, S.~E. 1999, \apj, 524, 262

\bibitem[{{Malesani} {et~al.}(2012){Malesani}, {Schulze}, {Kruehler}, {Fynbo},
  {Hjorth}, {Milvang-Jensen}, {Watson}, {de Ugarte Postigo}, {Tanvir},
  {Tagliaferri}, {Leloudas}, {Sollerman}, {Xu}, {Stritzinger}, \& {De
  Cia}}]{2012GCN..13277...1M}
{Malesani}, D., {et~al.} 2012, GRB Coordinates Network, 13277, 1

\bibitem[{{Melandri} {et~al.}(2012){Melandri}, {Pian}, {Ferrero}, {D'Elia},
  {Walker}, {Ghirlanda}, {Covino}, {Amati}, {D'Avanzo}, {Mazzali}, {Della
  Valle}, {Guidorzi}, {Antonelli}, {Bernardini}, {Bersier}, {Bufano},
  {Campana}, {Castro-Tirado}, {Chincarini}, {Deng}, {Filippenko}, {Fugazza},
  {Ghisellini}, {Kouveliotou}, {Maeda}, {Marconi}, {Masetti}, {Nomoto},
  {Palazzi}, {Patat}, {Piranomonte}, {Salvaterra}, {Saviane}, {Starling},
  {Tagliaferri}, {Tanaka}, \& {Vergani}}]{2012A&A...547A..82M}
{Melandri}, A., {et~al.} 2012, \aap, 547, A82

\bibitem[{{Metzger} {et~al.}(2010){Metzger}, {Mart{\'{\i}}nez-Pinedo},
  {Darbha}, {Quataert}, {Arcones}, {Kasen}, {Thomas}, {Nugent}, {Panov}, \&
  {Zinner}}]{2010MNRAS.406.2650M}
{Metzger}, B.~D., {et~al.} 2010, \mnras, 406, 2650

\bibitem[{{Nakar} \& {Sari}(2012)}]{2012ApJ...747...88N}
{Nakar}, E., \& {Sari}, R. 2012, \apj, 747, 88

\bibitem[{{Nomoto} {et~al.}(2003){Nomoto}, {Maeda}, {Mazzali}, {Umeda}, {Deng},
  \& {Iwamoto}}]{2003astro.ph..8136N}
{Nomoto}, K., {Maeda}, K., {Mazzali}, P.~A., {Umeda}, H., {Deng}, J., \&
  {Iwamoto}, K. 2003, arXiv:astro-ph/0308136

\bibitem[{{Ofek} {et~al.}(2007){Ofek}, {Cenko}, {Gal-Yam}, {Fox}, {Nakar},
  {Rau}, {Frail}, {Kulkarni}, {Price}, {Schmidt}, {Soderberg}, {Peterson},
  {Berger}, {Sharon}, {Shemmer}, {Penprase}, {Chevalier}, {Brown}, {Burrows},
  {Gehrels}, {Harrison}, {Holland}, {Mangano}, {McCarthy}, {Moon}, {Nousek},
  {Persson}, {Piran}, \& {Sari}}]{2007ApJ...662.1129O}
{Ofek}, E.~O., {et~al.} 2007, \apj, 662, 1129

\bibitem[{{Papish} \& {Soker}(2011)}]{2011MNRAS.416.1697P}
{Papish}, O., \& {Soker}, N. 2011, \mnras, 416, 1697

\bibitem[{{Pian} {et~al.}(2006){Pian}, {Mazzali}, {Masetti}, {Ferrero},
  {Klose}, {Palazzi}, {Ramirez-Ruiz}, {Woosley}, {Kouveliotou}, {Deng},
  {Filippenko}, {Foley}, {Fynbo}, {Kann}, {Li}, {Hjorth}, {Nomoto}, {Patat},
  {Sauer}, {Sollerman}, {Vreeswijk}, {Guenther}, {Levan}, {O'Brien}, {Tanvir},
  {Wijers}, {Dumas}, {Hainaut}, {Wong}, {Baade}, {Wang}, {Amati}, {Cappellaro},
  {Castro-Tirado}, {Ellison}, {Frontera}, {Fruchter}, {Greiner}, {Kawabata},
  {Ledoux}, {Maeda}, {M{\o}ller}, {Nicastro}, {Rol}, \&
  {Starling}}]{2006Natur.442.1011P}
{Pian}, E., {et~al.} 2006, \nat, 442, 1011

\bibitem[{{Pignata} {et~al.}(2011){Pignata}, {Stritzinger}, {Soderberg},
  {Mazzali}, {Phillips}, {Morrell}, {Anderson}, {Boldt}, {Campillay},
  {Contreras}, {Folatelli}, {F{\"o}rster}, {Gonz{\'a}lez}, {Hamuy},
  {Krzeminski}, {Maza}, {Roth}, {Salgado}, {Levesque}, {Rest}, {Crain},
  {Foster}, {Haislip}, {Ivarsen}, {LaCluyze}, {Nysewander}, \&
  {Reichart}}]{2011ApJ...728...14P}
{Pignata}, G., {et~al.} 2011, \apj, 728, 14

\bibitem[{{Podsiadlowski} {et~al.}(2004){Podsiadlowski}, {Mazzali}, {Nomoto},
  {Lazzati}, \& {Cappellaro}}]{2004ApJ...607L..17P}
{Podsiadlowski}, P., {Mazzali}, P.~A., {Nomoto}, K., {Lazzati}, D., \&
  {Cappellaro}, E. 2004, \apjl, 607, L17

\bibitem[{{Roberts} {et~al.}(2011){Roberts}, {Kasen}, {Lee}, \&
  {Ramirez-Ruiz}}]{2011ApJ...736L..21R}
{Roberts}, L.~F., {Kasen}, D., {Lee}, W.~H., \& {Ramirez-Ruiz}, E. 2011, \apjl,
  736, L21

\bibitem[{{Rosswog} \& {Ramirez-Ruiz}(2002)}]{2002MNRAS.336L...7R}
{Rosswog}, S., \& {Ramirez-Ruiz}, E. 2002, \mnras, 336, L7

\bibitem[{{Smith} \& {Owocki}(2006)}]{2006ApJ...645L..45S}
{Smith}, N., \& {Owocki}, S.~P. 2006, \apjl, 645, L45

\bibitem[{{Smith} {et~al.}(2012){Smith}, {Cenko}, {Butler}, {Bloom},
  {Kasliwal}, {Horesh}, {Kulkarni}, {Law}, {Nugent}, {Ofek}, {Poznanski},
  {Quimby}, {Sesar}, {Ben-Ami}, {Arcavi}, {Gal-Yam}, {Polishook}, {Xu},
  {Yaron}, {Frail}, \& {Sullivan}}]{2012MNRAS.420.1135S}
{Smith}, N., {et~al.} 2012, \mnras, 420, 1135

\bibitem[{{Soderberg} {et~al.}(2006){Soderberg}, {Kulkarni}, {Nakar}, {Berger},
  {Cameron}, {Fox}, {Frail}, {Gal-Yam}, {Sari}, {Cenko}, {Kasliwal},
  {Chevalier}, {Piran}, {Price}, {Schmidt}, {Pooley}, {Moon}, {Penprase},
  {Ofek}, {Rau}, {Gehrels}, {Nousek}, {Burrows}, {Persson}, \&
  {McCarthy}}]{2006Natur.442.1014S}
{Soderberg}, A.~M., {et~al.} 2006, \nat, 442, 1014

\bibitem[{{Soderberg} {et~al.}(2010){Soderberg}, {Chakraborti}, {Pignata},
  {Chevalier}, {Chandra}, {Ray}, {Wieringa}, {Copete}, {Chaplin},
  {Connaughton}, {Barthelmy}, {Bietenholz}, {Chugai}, {Stritzinger}, {Hamuy},
  {Fransson}, {Fox}, {Levesque}, {Grindlay}, {Challis}, {Foley}, {Kirshner},
  {Milne}, \& {Torres}}]{2010Natur.463..513S}
---. 2010, \nat, 463, 513

\bibitem[{{Sparre} {et~al.}(2011){Sparre}, {Sollerman}, {Fynbo}, {Malesani},
  {Goldoni}, {de Ugarte Postigo}, {Covino}, {D'Elia}, {Flores}, {Hammer},
  {Hjorth}, {Jakobsson}, {Kaper}, {Leloudas}, {Levan}, {Milvang-Jensen},
  {Schulze}, {Tagliaferri}, {Tanvir}, {Watson}, {Wiersema}, \&
  {Wijers}}]{2011ApJ...735L..24S}
{Sparre}, M., {et~al.} 2011, \apjl, 735, L24

\bibitem[{{Stanek} {et~al.}(2003){Stanek}, {Matheson}, {Garnavich}, {Martini},
  {Berlind}, {Caldwell}, {Challis}, {Brown}, {Schild}, {Krisciunas}, {Calkins},
  {Lee}, {Hathi}, {Jansen}, {Windhorst}, {Echevarria}, {Eisenstein}, {Pindor},
  {Olszewski}, {Harding}, {Holland}, \& {Bersier}}]{2003ApJ...591L..17S}
{Stanek}, K.~Z., {et~al.} 2003, \apjl, 591, L17

\bibitem[{{Stanek} {et~al.}(2005){Stanek}, {Garnavich}, {Nutzman}, {Hartman},
  {Garg}, {Adelberger}, {Berlind}, {Bonanos}, {Calkins}, {Challis}, {Gaudi},
  {Holman}, {Kirshner}, {McLeod}, {Osip}, {Pimenova}, {Reiprich}, {Romanishin},
  {Spahr}, {Tegler}, \& {Zhao}}]{2005ApJ...626L...5S}
---. 2005, \apjl, 626, L5

\bibitem[{{Troja} {et~al.}(2012){Troja}, {Sakamoto}, {Guidorzi}, {Norris},
  {Panaitescu}, {Kobayashi}, {Omodei}, {Brown}, {Burrows}, {Evans}, {Gehrels},
  {Marshall}, {Mawson}, {Melandri}, {Mundell}, {Oates}, {Pal'shin}, {Preece},
  {Racusin}, {Steele}, {Tanvir}, {Vasileiou}, {Wilson-Hodge}, \&
  {Yamaoka}}]{2012ApJ...761...50T}
{Troja}, E., {et~al.} 2012, \apj, 761, 50

\bibitem[{{Wiersema} {et~al.}(2012){Wiersema}, {Cucchiara}, {Levan},
  {Rapoport}, {Schmidt}, {Bersier}, {Perley}, {Cenko}, \&
  {Tanvir}}]{2012GCN..13276...1W}
{Wiersema}, K., {et~al.} 2012, GRB Coordinates Network, 13276, 1

\bibitem[{{Woosley} \& {Bloom}(2006)}]{2006ARA&A..44..507W}
{Woosley}, S.~E., \& {Bloom}, J.~S. 2006, \araa, 44, 507

\bibitem[{{Zhang} {et~al.}(2012){Zhang}, {Fan}, {Shen}, {Xu}, {Zhang}, {Wei},
  {Burrows}, {Zhang}, \& {Gehrels}}]{2012ApJ...756..190Z}
{Zhang}, B.-B., {et~al.} 2012, \apj, 756, 190

\end{thebibliography}
\end{document}